\documentclass[aps,prl,twocolumn,showpacs,superscriptaddress,groupedaddress]{revtex4}  
\usepackage{graphicx}  
\usepackage{dcolumn}   
\usepackage{bm}        
\usepackage{amssymb}   
\usepackage{lineno,hyperref}
\modulolinenumbers[3]

        \newcommand{\kbar}{$\mathrm {\bar{K}~}$}
        \newcommand{\Sz}{$\Sigma^0$}
          \newcommand{\Szs}{$\Sigma^0$}
        \newcommand{\Km}{K$^-$}
        \newcommand{\Lp}{$\Lambda$p}
        \newcommand{\Lg}{$\Lambda\gamma$}
        \newcommand{\Lpz}{$\Lambda\pi^0$p} 
        \newcommand{\Lstar}{$\Lambda(1405)$}
        \newcommand{\Szp}{$\Sigma^0$p}
        \newcommand{\ppk}{ppK$^-$}

\begin{document}

\widetext


\title{ \Km~absorption on two nucleons and \ppk~bound state search in the \Szp~final state}

\author
       {O.~V\'azquez~Doce$^{1,2}$,
        L.~Fabbietti$^{1,2}$,
        M.~Cargnelli$^3$, 
        C.~Curceanu$^4$, 
        J.~Marton$^3$,
        K.~Piscicchia$^{4,5}$,
        A.~Scordo$^4$, 
        D.~Sirghi$^4$, 
        I.~Tucakovic$^4$, 
        S.~Wycech$^6$,
        J.~Zmeskal$^3$,
        A.~Anastasi$^{4,7}$,
        F.~Curciarello$^{7,8,9}$,
        E.~Czerwinski$^{10}$,
        W.~Krzemien$^6$,
        G.~Mandaglio$^{7,11}$,
        M.~Martini$^{4,12}$,
        P.~Moskal$^{10}$,
        V.~Patera$^{13,14}$,
        E.~P\'erez~del~Rio$^4$ and
        M.~Silarski$^4$.
       \\~
       \\
       \textit{
       $^1$ Excellence Cluster 'Origin and Structure of the Universe', 85748 Garching (Germany)\\
       $^2$ Physik Department E12, Technische Universit\"at M\"unchen, 85748 Garching (Germany)\\
       $^3$ Stefan-Meyer-Institut f\"ur Subatomare Physik, 1090 Wien (Austria) \\
       $^4$ INFN, Laboratori Nazionali di Frascati, 00044 Frascati (Italy) \\
       $^5$ Museo Storico della Fisica e Centro Studi e Ricerche Enrico Fermi (Italy)\\
       $^6$ National Centre for Nuclear Research, 00681 Warsaw (Poland)\\
       $^7$ Dipartimento M.I.F.T. dell'Universit\`a di Messina, 98166 Messina (Italy)\\
       $^8$ Novosibirsk State University, 630090 Novosibirsk (Russia)\\
       $^9$ INFN Sezione Catania, 95129 Catania (Italy)\\
       $^{10}$ Institute of Physics, Jagiellonian University, 30-059 Cracow (Poland) \\
       $^{11}$ INFN Gruppo collegato di Messina, 98166 Messina (Italy) \\
       $^{12}$ Dipartimento di Scienze e Tecnologie applicate, Universit\`a 'Guglielmo Marconi', 00193 Roma (Italy)\\
       $^{13}$ Dipartimento di Scienze di Base e Applicate per l'Ingegneria, Universit\`a 'Sapienza', 00161 Roma (Italy)\\
       $^{14}$ INFN Sezione di Roma, 00185 Roma (Italy)
       }
       \\~
       }

\date{\today}

\begin{abstract}
We report the measurement of \Km~absorption processes in the \Szp~final state
and the first exclusive measurement of the  two nucleon absorption (2NA) with the KLOE detector. The 2NA process without further interactions 
 is found to be 12\% of the sum of all other
contributing processes, including absorption on three and more nucleons or 2NA followed by final state interactions
with the residual nucleons.
We also determine the possible contribution of the \ppk~bound state to the 
\Szp~final state. 
The yield of \ppk$ /\mathrm{K^-_{stop}}$ is found to be $(0.044 \pm 0.009\, stat ^{+ 0.004} _{- 0.005} \,syst)  \cdot 10^{-2}$
 but its statistical significance based on an F-test
is only 1$\sigma$.
\end{abstract}

\maketitle
\section{Introduction}

The study of the \kbar-nucleus interaction at low energies is of interest not only for quantifying the meson-baryon
potential with strange content \cite{Cab14}, but also because of its impact on models describing the structure of neutron stars  (NS) \cite{Nel}.
The \kbar-nucleus potential is attractive, as theory predicts \cite{Fuchs} and kaonic
atoms confirm \cite{sid:2011}, and this fact leads to the formulation of hypotheses 
about antikaon condensates
inside the dense interior of neutron stars.
Although recently measured heavy NS \cite{Dem10} constrain the equation of state of the latter
as being rather stiff and hence degrees of freedom other than neutrons are disfavoured and theoretical calculations about nuclear systems with high multiplicity of antikaons  present upper limits that disfavour the appearance of a condensate \cite{Gaz12}, experimental studies of the antikaon behaviour in nuclear matter are needed. 
The study of antikaons production in heavy-ion reactions at moderate energies ($E_{\mathrm{KIN}}\approx$ GeV),
with maximal reached baryon densities of $\rho\approx (3-4)\cdot \rho_0$ (with $\rho_0$ being the normal nuclear matter density)
has been carried out to find
evidence of a strong attractive potential between antikaons within dense nuclear matter \cite{KHI}.
However, the 
statistics collected so far \cite{fop15} does not allow for any conclusive statement about the role
played by kaons within dense nuclear matter. In this context
it is crucial that the theoretical models used to interpret the data 
properly include both the rather large cross-sections for antikaon absorption 
processes on nucleons and the presence of the \Lstar~resonance \cite{Cab14}.
 Indeed, an antikaon produced within nuclear matter can undergo absorption upon one or more nucleons and the
 measurement of such processes is not yet exhaustive, even at normal nuclear densities \cite{KATZ,KEK}.
  Absorption processes also play an important role in the understanding of kaonic atoms, where 
 a substantial multi-nucleon component is put forward by some theoretical models \cite{Gal13}. 
 The \Lstar~link to the antikaon-nucleon interaction resides in the fact that
 theory describes this resonance as generated
 dynamically from the coupling of the \kbar-p and the $\Sigma$-$\pi$ channels \cite{theoL1405}.
  Hence the \Lstar~can be seen, at least partially, as a \kbar-p bound state.
 Despite of several experimental measurements \cite{expL1405}, not even the vacuum properties of the \Lstar~are yet pinned down 
 precisely and those can also be modified at finite baryonic densities, with major implications for the  \kbar dynamics in the medium.\\
Following the line of thought employed to interpret the \Lstar, one or more nucleons could be kept together 
by the strong attractive interaction between antikaons and nucleons, and then so-called kaonic bound states 
as pp\Km~or ppn\Km~might be formed.
The observation of such states and the measurement of their binding energies and widths would provide a quantitative
measurement of the \kbar-nucleon interaction in vacuum, providing an important reference for the investigation of the in-medium
properties of $\mathrm{\bar{K}}$.
For the di-baryonic kaonic bound state \ppk, theoretical predictions deliver a wide range of binding energies and widths
\cite{theoppK} and experimental results are contradictory \cite{ppKExp}.
For the search of such states in \Km-absorption experiments, the competing multi-nucleonic absorption plays a
fundamental role.\\
This work focuses on the analysis of the \Szp~final state produced in absorption processes of 
\Km~on two or more nucleons and the search for a signature of the $ppK^- \rightarrow \Sigma^0 +p$ kaonic bound state.
The chosen \Szp~final state is free from the ambiguities present in the analysis of  the $\Lambda$p state
 considered in previous works \cite{KEK}.
Moreover, this study represents the first attempt of combining a quantitative understanding of the absorption processes
and contributing background sources with the test of different hypotheses for the \ppk~bound state properties. 
\section{$\Sigma^0$p Selection and Interpretation}
The analysed data corresponds to a total integrated luminosity of $1.74 fb^{-1}$ collected in 2004-2005 with the KLOE detector \cite{KLOE} located at the DA$\Phi$NE 
e$^+$e$^-$ collider \cite{dafne}. There, $\phi$ mesons are produced nearly at rest, providing an almost monochromatic source of \Km~with a momentum of $\sim \,127$ MeV/c.\\
The data here presented was taken by the KLOE collaboration and provided to the authors for an
independent analysis.
The KLOE detector consists of a large acceptance cylindrical drift chamber (DC) of $3$~m length and $2$~m radius surrounded 
by an electromagnetic calorimeter (EMC) inside an axial magnetic field of $0.52$~T.
The DC provides a spatial resolution of 150~$\mu$m and $2$~mm in the radial and longitudinal coordinates, respectively,
and a transverse momentum resolution of $\sigma_{p_T}/p_T \sim$ 0.4\% for large angle tracks.
The EMC is composed of barrel and end-cap modules covering 98\% of the solid
angle with energy and time resolutions of $\sigma_{E}/E$=~5.7\%/$\sqrt{E(GeV)}$
and $\sigma_t$=~54 ps/ $\sqrt{E(GeV)}$, respectively.
The DC entrance wall is composed of $750\,\mu$m carbon fibre with inner and outer layers of aluminium of 100~$\mu$m
thickness. The number of stopped \Km~in this wall is calculated by combining the experimental K$^+$ tagging efficiency, the 
luminosity information and a Monte Carlo simulation to determine the rate of \Km~stopped in the DC wall. 
The decay nearly at rest of the $\phi$ meson allows to tag \Km events by the identification of a K$^+$ track in the opposite hemisphere of the DC.
The extracted total number of stopped 
\Km~is equal to $(3.25\pm0.06)\cdot 10^8$. This value is used to normalize the measured yields of the different absorption processes.
Both in flight and at rest \Km~absorptions can occur and a weight of 50\% is assigned to each process for the normalization. \\
The starting point for the selection of \Km~absorption processes leading to $\Sigma^0$p final state is the identification of a $\Lambda(1116)$ hyperon
through its decay into protons and negative pions (BR = 63.8\%).
Proton and pion track candidates are selected via dE/dx measurement in the DC.
For each proton and pion candidate a minimum track length of at least 30 and 50 cm is required, respectively.
The track length must also be larger than 50\% of the expected length value
calculated by extrapolating
the measured momentum at the DC entrance. Additionally, proton candidates must have a momentum higher than $170$ MeV/c.
These selections aim to improve the purity of the particle identification, minimize the pion contamination in the proton sample and
minimize the contribution from low momentum tracks that are emitted parallel to the DC wires and reach the EMC barrel.
The reconstructed $M_{p\pi^-}$ invariant mass shows a mean value of $1115.753 \pm 0.002$~MeV/c$^2$ for the mass,
with a resolution of $\sigma$=~$0.5$~MeV/c$^2$, well in agreement with the PDG value \cite{PDG}. 
The $\Lambda$ candidates are selected using the following cut:
$1112 <\, M_{p \pi^-} < \,1118$~MeV/c$^2$.\\
\begin{figure}[ht]
\includegraphics*[height=6.2cm,width=8.5cm]{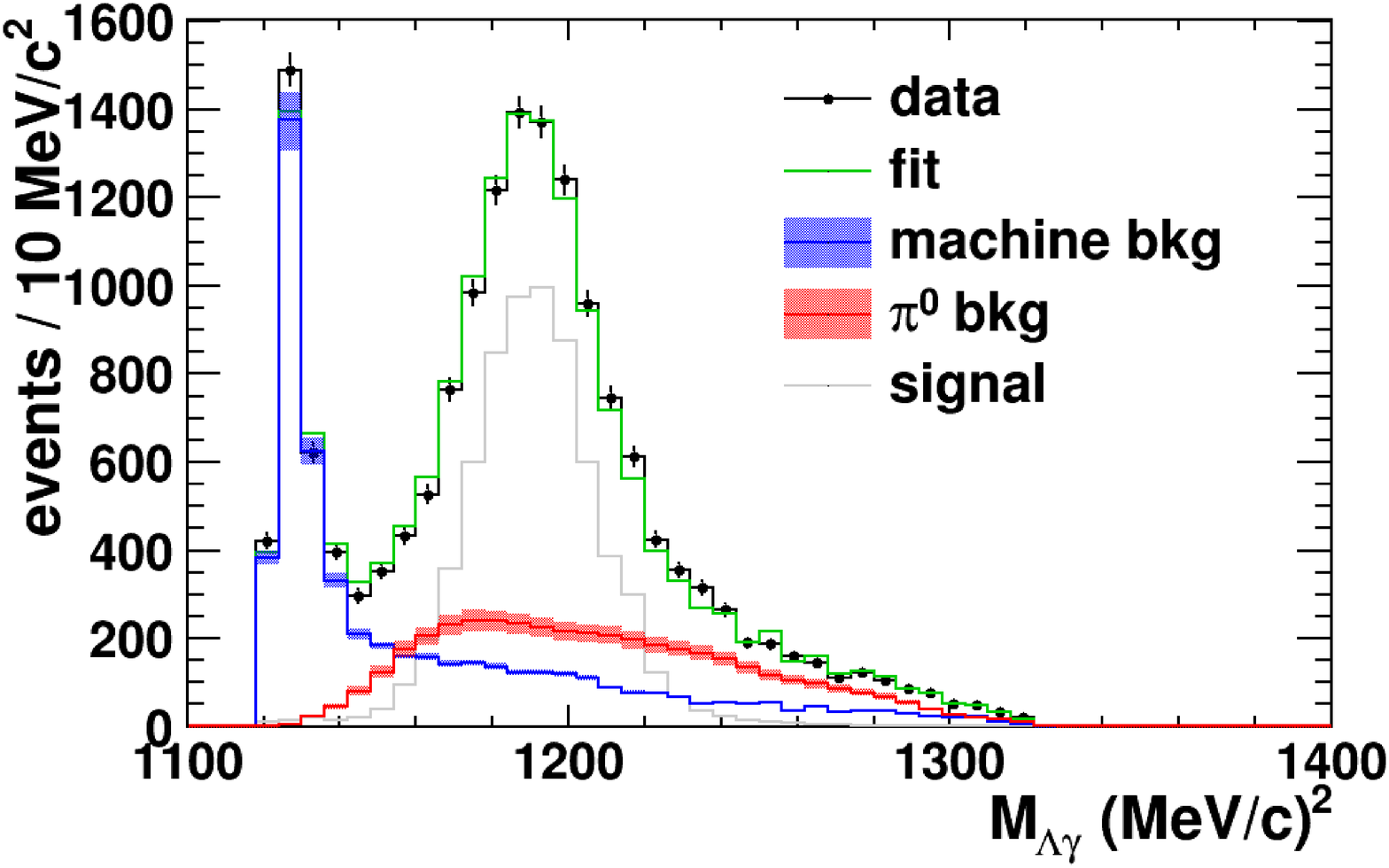}
\caption{\label{s0mass} (Color online) \Lg~invariant mass distribution. The black symbols represent the experimental data, the blue and the
red histograms are the contribution from the machine background and events that contain a \Lpz~in the final state,
respectively. The gray histogram shows the simulated \Sz~signal and the green one the overall fit to the data (see text for details).}
\end{figure}
A common vertex between the $\Lambda$ candidate and an additional proton 
track 
is then searched for.
The obtained resolution on the radial coordinate ($\rho_{\Lambda p}$) for the \Lp~vertex is $12$~mm,
and this topological variable is used to select the \Km~absorption processes inside the DC wall.
The \Lp~invariant mass resolution is evaluated with a 
phase space Monte Carlo simulation where the proton and $\Lambda$ momenta are varied from $100$ to $700$~MeV/c and 
is found to be equal to $1.1$~MeV/c$^2$.
The contamination to the proton sample for the $\Lambda p$ 
final state due to heavier particles (deuterons or tritons) is estimated to be less than 2\% by MC simulations.\\
The \Sz~candidates are identified through their decay into \Lg~pairs.
After the reconstruction of a \Lp~pair, the photon selection is carried out via its identification in the EMC.
Photon candidates are selected by applying a cut on the difference between
 the EMC time measurement and the expected time of arrival
of the photon within $-1.2 < \Delta t < 1.8$~ns.
The resulting \Lg~invariant mass distribution is shown in Fig.~\ref{s0mass}, where the \Sz
signal is visible above a background distribution. \\
The following kinematic distributions are considered simultaneously 
in a global fit to extract the contributions of the different absorption processes:
 the \Szp~invariant mass, the cos($\theta_{\Sigma^0 p}$), 
the \Sz~and the proton momenta. 
The processes that are taken into account in the fit 
of the experimental data are:
\begin{enumerate}
\item \label{reaction1} $K^-$A $\rightarrow$ \Szs-($\pi$)~$p_{\mathrm{spec}}$ (A'), 
\item \label{reaction2} $K^-$pp $\rightarrow$ \Szs-p (2NA),
\item \label{reaction3} $K^-$ppn $\rightarrow$ \Szs-p-n (3NA),
\item \label{reaction4} $K^-$ppnn $\rightarrow$ \Szs-p-n-n (4NA).
\end{enumerate}
This list includes the \Km~absorption on two nucleons with and without final state interaction for the
\Szp~state and processes involving more than two nucleons in the initial state. 
These contributions are either extracted from experimental data samples or
modelled via simulations and digitised.
Nevertheless, the background contributions must be determined and subtracted prior to the global fit.\\
Two kinds of background contribute to the analysed \Szp~final state: 
the machine background and the events with \Lpz~in the final state. Both are quantified using experimental data. 
The machine background originates from spurious hits in the EMC that enter the photon time coincidence window.
It is emulated by a side band analysis, selecting events with EMC hits outside the coincidence window
 ($-4<\Delta t<-2$~ns and $3<\Delta t<8.2$~ns).
The \Lpz~background originates from single nucleon absorptions of \Km~leading to the creation of a $\Sigma \pi^0$ pair.
The $\Sigma$ hyperon is then undergoing an internal conversion process on a residual nucleon ($\Sigma N \rightarrow \Lambda N$)
leading to a \Lpz~final state.
Events with two photon candidates within the selected time window and with a $\gamma\gamma$ invariant 
mass around  the $\pi^0$ nominal mass (3$\sigma$ of the experimental resolution of $17$~MeV/c$^2$) 
are selected to emulate the \Lpz~background.
Both experimental samples are used together with a simulation of the \Sz~signal to fit the $\Lambda$-photon 
invariant mass distribution in order to extract the machine and \Lpz~background contributions.
Full scale simulations of the \Sz~reconstruction in the \Lg~channel lead to a mass mean value and $\sigma$
of 
$1189$ and $14.5$~MeV/c$^2$, respectively. The mean value is slightly shifted with respect to the \Sz~nominal
mass reported in \cite{PDG} because of the small bias introduced by the reconstruction. This simulation is used to 
fit the \Sz~signal on the \Lg~ distribution shown in Fig.~\ref{s0mass}.\\
Figure~\ref{s0mass} shows the results of the fit to the \Lg~invariant mass, 
with the background and signal components. The black symbols refer to the experimental data,
the blue histogram to the fitted machine background, the red histogram to the \Lpz~background,
the gray one to the simulated \Sz~signal and the green histogram to the sum of all fit contributions.
The errors shown for the background distributions
represent the statistical errors of the fit.
To enhance the purity of the experimental data for the following analysis steps, a cut on the \Lg~invariant mass
 around the nominal \Sz~mass is applied. The applied cut, $1150< M_{\Lambda\gamma}<1235$~MeV/c$^2$, 
 corresponds to $3\sigma$ of the experimental resolution, and is verified with MC simulation.
 The contribution of the machine and \Lpz~backgrounds within the selected \Lg~invariant mass is
  (14.6$\pm$0.8)\% and (26.1$\pm$2.7)\%, respectively.
The machine background is directly subtracted from the experimental data for each kinematic distributions
used for the global fit. 
The fit error is added to the statistical errors of the experimental data after the background subtraction.
The \Lpz~background is considered in the global fit using the obtained yield as a starting value.
\section{Determination of the Absorption Processes} 
\begin{figure*}
\includegraphics*[height=13.cm,width=16.5cm]{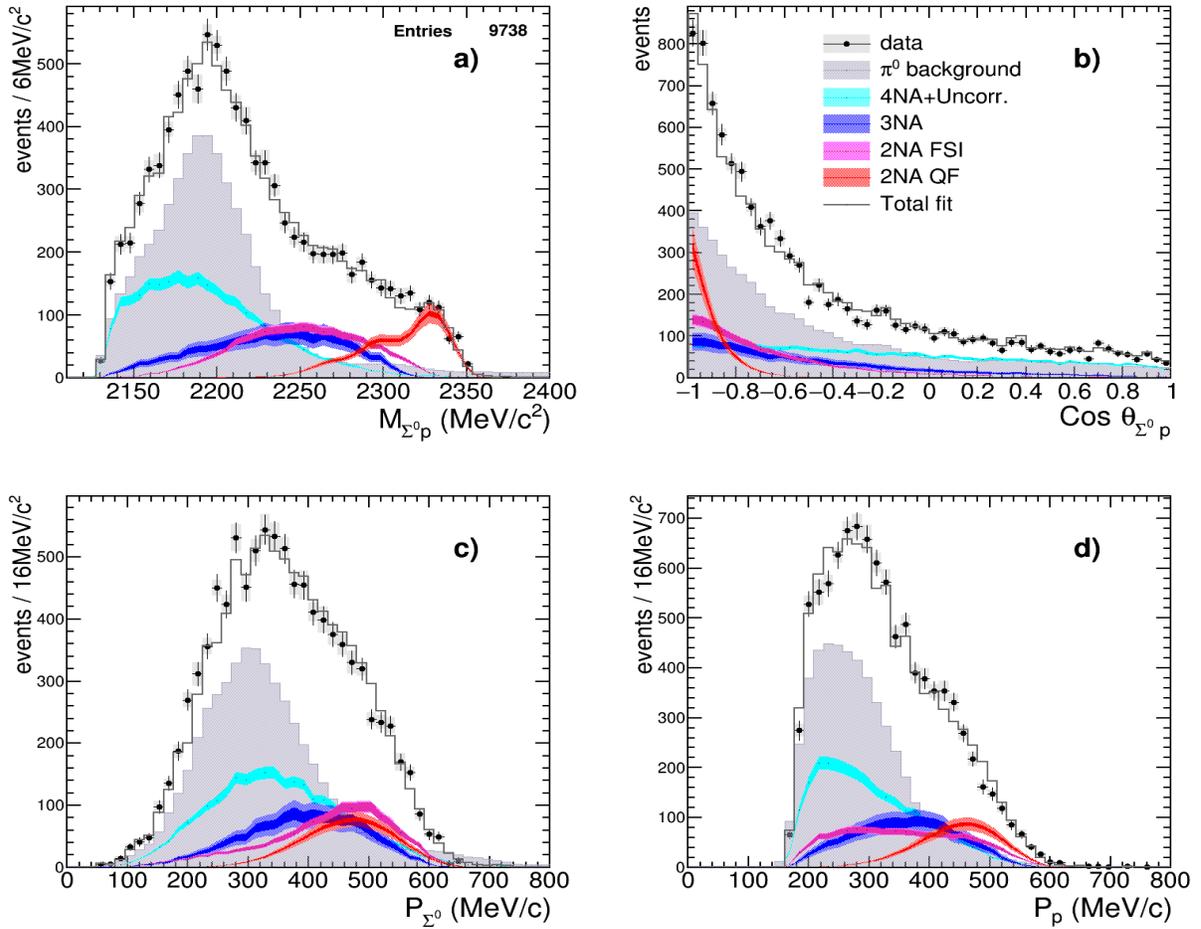}
\caption{\label{fit} (Color online) Experimental distributions of the $\Sigma^0$p invariant mass, cos($\theta_{\Sigma^0 p}$),
 $\Sigma^0$ momentum and proton momentum 
together with the results of the global fit. The experimental data after the subtraction of the machine background 
 are shown by the black circles, the systematic errors are represented by the boxes and the coloured histograms
  correspond to the fitted signal distributions where the light-coloured bands show the fit errors and the darker
   bands represent the symmetrised systematic errors. The gray line show the total fit distributions (see text for details).}
\end{figure*}
The cocktail of processes considered for the global fit is obtained as follows.
Process \ref{reaction1} corresponds to the uncorrelated production of a proton from the fragmentation of the residual nucleus 
together with a \Sz~production from the \Km~absorption. 
This contribution is obtained from experimental data containing  $\Lambda$-triton-proton, 
$\Lambda$-deuteron-proton or $\Lambda$-proton-proton in the final state
to emulate the case where the selected proton is barely correlated with the $\Lambda$. \\
For the simulation of the absorption processes \ref{reaction2}-\ref{reaction4} a $^{12}$C target is considered. 
The Fermi momentum of the interacting nucleons inside the $^{12}$C,
 the initial momentum of the absorbed \Km~and the mass difference between the initial and residual nuclei are 
 used in the calculation of the event kinematic.
The Fermi momentum distribution of the nucleons in $^{27}$Al target is only 9\% higher in comparison with $^{12}$C.
  The mass difference of the initial and residual nuclei varies only by 0.3\% when considering $^{27}$Al
  in comparison with $^{12}$C and this value is lower than the experimental resolution of the \Szp~invariant mass.
 For all the considered reactions, the emitted nucleons are required to have a total momentum above the $^{12}$C
 Fermi momentum to be able to leave the nucleus.\\
 For the 2NA (process \ref{reaction2}), two cases are studied. One including the final state interaction (2NA-FSI) of the \Sz~or proton 
with the residual nucleus, and the second assuming no FSI at all (2NA-QF). In the case of the FSI-free production of the 
\Szp~pair, only the fragmentation of the residual nucleus is considered. This results in a lower \Szp~
invariant mass leaving all other kinematic variables unaffected. 
 The FSI for the \Sz~and p is implemented by allowing the
 outgoing proton or \Sz~to scatter with the residual nucleons. The nucleon 
 momentum is sampled according to the Fermi distribution and the scattering probability is assumed to be equal to 
 50\% for both the cases of one and two collisions. A more sophisticated propagation of the hit probability was 
 investigated \cite{oset}, but the results show no major differences in the resulting kinematic distributions. 
 This motivates the simplification. \\
The processes \ref{reaction1}-\ref{reaction4} together with the \Lpz~background sample are used for the global fit.
The starting value for the \Lpz~yield is extracted from the fit to the \Lg~invariant mass 
but this component is free to vary within $2\sigma$ in the global fit.
Panels (a-d) in Fig.~\ref{fit} show the
experimental distributions for the \Szp~invariant mass, the cos($\theta_{\Sigma^0 p}$), 
and the \Sz~and proton momenta. 
together with the fit results.
The black points represent the experimental data after the subtraction of the 
machine background with the systematic errors shown by boxes.
The gray filled histogram represents the \Lpz~background. The cyan distributions show the sum of the
4NA simulation together with the uncorrelated production of the \Szp~final state. The blue distribution
represents the 3NA and the magenta histogram the 2NA-FSI. The red distribution shows
the 2NA-QF. The gray line is the resulting total fit. For each fitted distribution the light error
band corresponds to the statistical error resulting from the fit, while the darker band visualises the
symmetrized systematic error.\\
The systematic errors for the experimental and simulated distributions are obtained by
varying the minimum momentum required for the proton track selection, the time window for
 the selection of signal and machine background, the yields of the machine background distributions and
the selection of the \Sz~mass. The obtained resolution for the \Lg~invariant mass and \Sz~momentum
is equal to 3.5~MeV/c$^2$ and 6~MeV/c, respectively. \\
The minimum momentum for the proton tracks is varied within $10$~MeV/c of the central value
 of $170$~MeV/c. Variations of 15\% are tested for the time windows used to select the machine
  background, the photon signal and $\pi^0$ background selection independently.
As for the machine background subtraction, the systematic error is evaluated by repeating the 
fit allowing variations within 1$\sigma$ of the initial \Sz~mass fit parameters.
For what concerns the simulated distributions, the systematic errors are also evaluated for the minimum momentum
of the nucleons required to exit the nucleus in the absorption simulation and the probability of having more than one 
collision when simulating the FSI with the residual nucleus following a 2NA process.\\
The minimum momentum for the nucleons is sampled according to the Fermi momentum distribution
 between $170$ and $220$~MeV/c. The systematic error is evaluated by varying the two
  boundaries by 15\% in both directions.
For the systematic variation of the probability of having one or two collisions for the FSI process two cases, 
40/60\% and 60/40\%, are evaluated respectively. \\
The final fit results deliver the contribution of the different channels to the analysed 
\Szp~final state. The best fit delivers a $\chi^2/ndf$ of $0.85$.
The emission rates extracted from the fit are normalised to the total number of stopped antikaons,
 as summarised in table 1. 
\begin{table}[h]
\renewcommand{\arraystretch}{1.3}
\begin{tabular}{|l|l|l|l|}
\hline
          & yield / K$^-_{stop} \cdot 10^{-2}$ & $\sigma_{stat} \cdot 10^{-2}$ & $\sigma_{syst} \cdot 10^{-2}$ \\ \hline
2NA-QF    & ~~~0.127 & ~$\pm$ 0.019 &  ~~$^{+0.004} _{-0.008}$  \\ \hline
2NA-FSI   & ~~~0.272 & ~$\pm$ 0.028 &  ~~$^{+0.022}_{ -0.023}$   \\ \hline
Tot 2NA   & ~~~0.376 & ~$\pm$ 0.033 & ~~$^{+0.023}_{ -0.032}$     \\ \hline
3NA       & ~~~0.274 & ~$\pm$ 0.069 &  ~~$^{+0.044}_{ -0.021}$     \\ \hline
Tot 3 body & ~~~0.546 & ~$\pm$ 0.074 & ~~$^{+0.048}_{ -0.033}$    \\ \hline
4NA + bkg. & ~~~0.773 & ~$\pm$ 0.053  & ~~$^{+0.025}_{ -0.076}$     \\ \hline
\end{tabular}
\caption{Production probability of the \Szp~final state for different intermediate processes
 normalised to the number of stopped \Km~in the DC wall. The statistical and systematic errors are shown as well.}
\label{table1}
\end{table}
The fit results lead to the first measurements of the genuine 2NA-QF for the final state \Szp~in
 reactions of stopped \Km~on targets of $^{12}$C 
and $^{27}$Al. This contribution is found to be only 12\% of the total absorption cross-section.

Even if the employed simulation model is rather simplified, the treatment of 
2NA-QF is
satisfactory for our purpose
and the signature of this component well distinguishable from the other contributions,
especially in the \Szp~invariant mass distribution.
On the other hand, a clear disentanglement of the 3NA process from the 2NA followed by FSI is 
difficult, due to the overlap of the
relevant kinematic variables over a wide range of the phase space.  
Two tests were performed that demonstrate that both physical processes 
should be included in the fit. First, if the 3NA contribution is switched off a variation of the reduced $\chi^2$ of 0.19 from 0.85 (the best 
fit) to 1.05 is observed. 
 Such effect is mainly due to the fact that the $\Sigma^0$ and the proton momentum distributions are no longer well described.
The other kinematic distributions are less sensitive to this contribution.
In particular, the $\chi^2$ calculated for the fit result of the proton momentum distribution only is deteriorated by 47\%
when excluding the 3NA contribution from the fit. As a second limiting case the 2NA + FSI contribution was discarded, leading to a reduced $\chi^2$ of 1.18. In this case the cos($\theta_{\Sigma^0 p}$) and \Szp~invariant mass distributions are not properly reproduced. \\
The uncorrelated emission of the \Szp~is also not distinguishable from 
the 4NA process and hence these two contributions are added up. 
\section{Search for the \ppk~bound state signal }
\label{secBS}
\begin{figure*}
\includegraphics*[height=6.6cm,width=15.5cm]{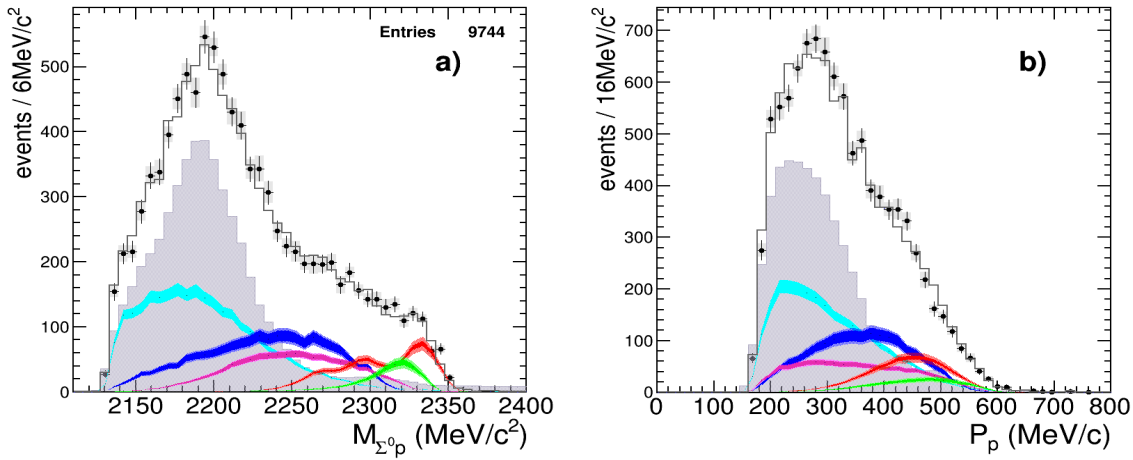}
\caption{\label{bs}(Color online) \Szp~invariant mass and proton momentum distributions together with the results
of the global fit including the \ppk. The different contributions are labeled as in Fig.~\ref{fit} and the green
histograms represent the \ppk~signal.
}
\end{figure*}

\begin{figure}
\includegraphics*[height=6.2cm,width=8.5cm]{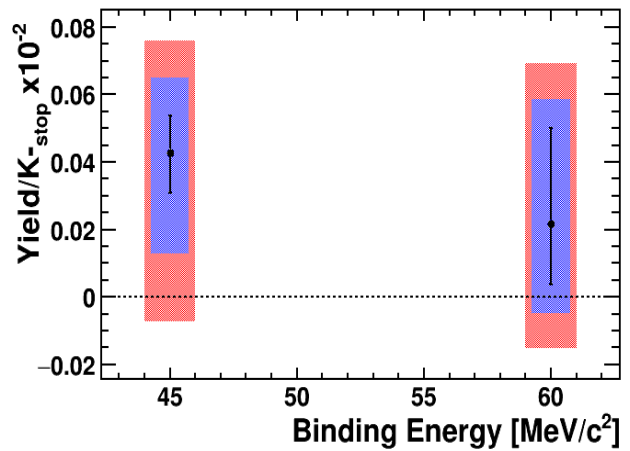}
\caption{\label{bssigmas} (Color online) \ppk~yield normalised to the number of stopped \Km~for the two best fits corresponding to binding
energies of $45$ and $60$~MeV/c$^2$ and width of $30$~MeV/c$^2$ for the \ppk~bound state. The errors are only statistical calculated
by MINOS at 1 (black line), 2 (blue boxes) and 3 (red boxes) $\sigma$.}
\end{figure}

\begin{figure}
\includegraphics*[height=6.2cm,width=8.5cm]{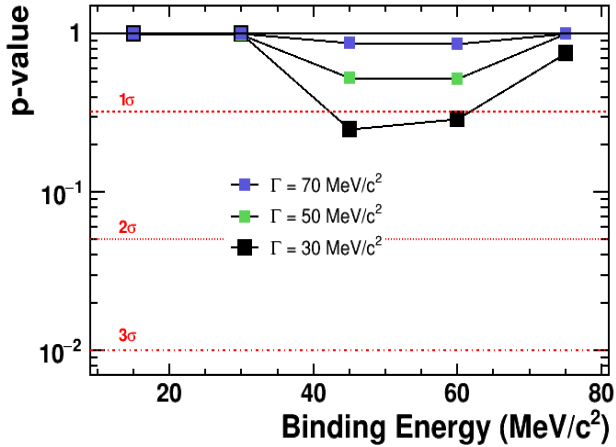}
\caption{\label{ftest}(Color online) p-value resulting from the F-test that compares the two
fitting hypotheses. Horizontal lines showing up to 3$\sigma$ are drawn.}
\end{figure}

The last step of the analysis consists in the search of the \ppk~bound state produced in \Km~interactions with nuclear targets, decaying into a $\Sigma^0$p pair. 
The \ppk~are simulated similarly to the 2NA-QF process but sampling the mass of the \ppk~state
with a Breit-Wigner distribution, rather than the Fermi momenta of the two nucleons in the initial state.
The event kinematic is implemented by imposing the momentum conservation of the \ppk-residual nucleus
system.
Different values for the binding energy and width varying within $15-75$~MeV/c$^2$ and $30-70$~MeV/c$^2$
 in steps of $15$ and $20$~MeV/c$^2$, respectively, are tested. This range is selected according to theoretical predictions \cite{theoppK} and taking into account the experimental resolution.
The global fit is repeated adding the \ppk~state to the processes \ref{reaction1}-\ref{reaction4}. 
The best fit ($\chi^2$/ndf=$\,0.807$) is obtained for a \ppk~candidate with a binding energy of $45$~MeV/c$^2$ and a width
 of $30$~MeV/c$^2$, respectively. 
Figure \ref{bs} shows the results of the best fit for the \Szp~invariant mass and proton momentum 
distributions where the \ppk~bound state contribution is shown in green. The resulting  yield
normalised to the number
of stopped \Km~is \ppk$ /\mathrm{K^-_{stop}} =  (0.044 \pm 0.009 stat ^{+ 0.004} _{- 0.005} syst)  \cdot 10^{-2}$.
Figure~\ref{bssigmas} shows the yield results from the two best fits of the bound state with a width of $30$~MeV/c$^2$ and a 
binding energy of $45$ and $60$~MeV/c$^2$, respectively, with statistical errors calculated
by MINOS at 1$\sigma$ (black line), 2$\sigma$ (blue boxes) and 3$\sigma$ (red boxes).
The inclusion of the \ppk~bound state to the global fit introduces an additional parameter
 and this improves 
the fit quality. Considering also
that the improvement of the $\chi^2$ is only marginal, a F-Test is carried out
to compare the two models: with and without \ppk~bound state. 
This test consists in evaluating the statistical significance of the model with the \ppk,
 accounting for the additional fit parameter, by comparing the residuals and number 
 of degrees of freedom of two models.
The resulting $F$ value reads as follows:
\begin{equation}
F = \frac{(SSE_1 - SSE_2) / (ndf_1 - ndf_2) }{ SSE_2 / ndf_2}
\label{yield}
\end{equation}
with $SSE$ being the quadratic sum of the residuals bin per bin and $ndf$ the number of degrees of 
freedom of each model.
The global p-value associated to the obtained $F$ value and from the number of parameters in each 
model is shown in Fig.~\ref{ftest} for bound state simulations 
with a width of $30$, $50$ and $70$~MeV/c$^2$ as a function of the binding energy. 
One can see that even the best fit corresponds to a p-value equal to $0.25$ and hence to a 
significance of $1~\sigma$.\\
This implies that although the fit favours the presence of an additional component that 
can be parametrised with a Breit-Wigner distribution with a certain mass and width, its significance is 
not sufficient to claim the 
observation 
of the bound state.
\section{Summary}
\label{summ}
We have presented the analysis of the \Km~absorption processes leading to the $\Sigma^0$p
 final state measured with the KLOE detector.
It is shown that the full kinematics of this final state can be reconstructed and 
a global fit of the kinematic variables allows to pin down quantitatively the different contributing processes.
A cocktail of processes including simulations of the \Km~absorption on two or more nucleons with or without
final state interactions and background processes estimated with
experimental data is used for the global fit.
The absorption on two nucleons without final state interaction (2NA-QF) is isolated
 and the yield normalised to the number of absorbed \Km~is presented in this
work for the first time. 
 It is shown that it is difficult to distinguish between the case where \Km~are 
 absorbed on three nucleons (3NA) or when the two-nucleon absorption is followed by a final 
 state interaction (2NA-FSI). For this purpose the data should be further interpreted 
 with the help of theoretical calculations.
 The 2NA-QF yield is found to be about 20\% of the sum of the 3NA and 2NA-FSI processes.
 If one considers the ratio of the 2NA-QF to all other simulated processes a value of about 12 \% is obtained. 
 Hence, we conclude that the contribution of the 2NA-QF processes for \Km~momenta lower
  than $120$~MeV/c is much smaller in comparison with other processes. \\
 A second fit of the experimental data including the contribution of a \ppk~bound state 
 decaying into a \Szp~final state is carried out. 
 A systematic scan of possible binding energies and widths varying within $15-75$~MeV/c$^2$ and $30-70$~MeV/c$^2$, respectively, 
 is performed and the best value of the total reduced $\chi^2$ is achieved for the hypothesis of a \ppk~with a binding energy of $45$~MeV/c$^2$ and a width of $30$~MeV/c$^2$. The corresponding \ppk~yield extracted from the fit is 
\ppk$ /\mathrm{K^-_{stop}} = (0.044 \pm 0.009 stat ^{+ 0.004} _{- 0.005} syst)  \cdot 10^{-2}$. A F-test is conducted
 to compare the simulation models with and without the \ppk~signal and to extract the significance of the result. 
 A significance of only 1 $\sigma$ is obtained for the \ppk~yield 
  result.
This shows that although the measured spectra are compatible with the
hypothesis of a contribution of the channel $ppK^- \rightarrow \Sigma^0 +p$, the significance
 of the result is not sufficient to claim the 
observation
of this state.

\section{Acknowledgments}
\label{Acknowledgments}
We acknowledge the KLOE collaboration for their support and for having provided us the data and the tools to perform the anaysis presented in this paper.

\end{document}